\documentclass[10pt,journal,compsoc]{IEEEtran}
\usepackage{amsmath,amsfonts}
\usepackage{algorithmic}
\usepackage{algorithm}
\usepackage{array}
\usepackage[caption=false,font=normalsize,labelfont=sf,textfont=sf]{subfig}
\usepackage{textcomp}
\usepackage{stfloats}
\usepackage{url}
\usepackage{verbatim}
\usepackage{graphicx}
\usepackage{cite}
\usepackage{fontawesome5}
\usepackage{ragged2e}
\usepackage{xcolor}
\usepackage[table]{xcolor} 
\usepackage{booktabs} 
\usepackage{ragged2e}
\newcolumntype{P}[1]{>{\RaggedRight\hspace{0pt}}p{#1}}
\definecolor{tableGray}{RGB}{243, 244, 245}
\definecolor{borderGray}{RGB}{229, 230, 233}
\usepackage{hyperref}
\usepackage{enumitem}

\hypersetup{
    colorlinks=true,
    linkcolor=black,
    filecolor=black,      
    urlcolor=black,
    citecolor=black,
}
\usepackage[many]{tcolorbox} 
\definecolor{tableGray}{RGB}{243, 244, 245}
\definecolor{borderGray}{RGB}{229, 230, 233}

\newtcolorbox{boxA}{
    colback = tableGray, 
    boxrule = 0pt  
}

\def\blackbar#1#2{
   {\color{black}\rule{#1mm}{4pt}}  #2}

\begin{document}


\title{Manifestations of Empathy in Software Engineering: How, Why, and When It Matters}


\author{Hashini~Gunatilake,~\IEEEmembership{Graduate Student Member,~IEEE},
        John~Grundy~\IEEEmembership{Fellow,~IEEE},
        Rashina~Hoda~\IEEEmembership{Member,~IEEE},
        Ingo~Mueller
\IEEEcompsocitemizethanks{\IEEEcompsocthanksitem Gunatilake, Grundy, and Hoda are with Faculty of IT, Monash University, Melbourne, Australia \protect\\
Mueller is with Monash Centre for Health Research \& Implementation, Monash Health, Melbourne, Australia \protect\\

Contact E-mail: hashini.gunatilake@monash.edu}
\thanks{Manuscript received April 23, 2025; revised August 8, 2023.}}

\markboth{Submitted to IEEE Transactions on Software Engineering}%
{Gunatilake \MakeLowercase{\textit{et al.}}: Manifestations of Empathy in Software Engineering}


\maketitle

\begin{abstract}
Empathy plays a crucial role in software engineering (SE), influencing collaboration, communication, and decision-making. While prior research has highlighted the importance of empathy in SE, there is limited understanding of how empathy manifests in SE practice, what motivates SE practitioners to demonstrate empathy, and the factors that influence empathy in SE work. Our study explores these aspects through 22 interviews and a large scale survey with 116 software practitioners. 
Our findings provide insights into the expression of empathy in SE, the drivers behind empathetic practices, SE activities where empathy is perceived as useful or less relevant, and the other factors that influence empathy. In addition, we offer practical implications for SE practitioners and researchers, offering a deeper understanding of how to effectively integrate empathy into SE processes.
\end{abstract}


\section{Introduction} \label{sec:introduction}
Empathy is ``the ability to experience the affective and cognitive states of another person, while maintaining a distinct self, in order to understand the other'' \cite{guthridge2021taxonomy}. In software engineering (SE), empathy plays a critical role in shaping collaboration, communication, and overall team dynamics, as well as it has been linked to improved mental health and well-being among software practitioners \cite{cerqueira2024empathy, gunatilake2025theory}. Recent SE research has focused on examining applicability of empathy through various models \cite{gunatilake2023empathy}, identifying enablers and barriers \cite{gunatilake2024enablers}, and exploring its meaning, significance, and practices from practitioners' perspectives \cite{cerqueira2023thematic}. 
In a previous study, we developed a theory explaining the role of empathy on the interactions between developers and stakeholders \cite{gunatilake2025theory}. We also presented actionable guidelines for practitioners to enhance empathy in professional settings \cite{gunatilake2025guidelines}. 

\textcolor{black}{However, while these prior studies have explored various aspects of empathy in SE, they do not provide a comprehensive understanding of its manifestations and conceptualisations, the ways in which empathy is perceived, expressed, and demonstrated in SE practices. Previous research tends to focus on specific elements of empathy, such as its impact \cite{gunatilake2025theory, cerqueira2023thematic, cerqueira2024empathy}, the causes of empathy \cite{gunatilake2025theory}, strategies for addressing empathy barriers \cite{gunatilake2024enablers}, and guidelines to enhance empathy adaptation \cite{gunatilake2025guidelines}. We found only one study that examined the meaning of empathy for software practitioners, which also identified a few SE activities where empathy is considered useful, such as coding, code reviews, and testing \cite{cerqueira2023thematic}. This study derived its findings from grey literature, particularly blog posts from developers,\footnote{https://dev.to} providing valuable insights into how empathy is perceived within this specific community. However, exploring why software practitioners are motivated to apply empathy in their work, whether they distinguish between a general definition of empathy and its meaning within SE, what broad SE activities they consider empathy to be beneficial or less relevant, and other factors influencing empathy remain unexplained in SE literature. We argue that addressing these gaps is essential for developing a comprehensive understanding of how empathy manifests in SE.}


The study presented in this paper investigates how empathy manifests in SE by examining practitioners' perspectives through a multi-method research study involving an interview study and a qualitative survey. Specifically, we explore how practitioners define empathy in general and in SE contexts, their motivations for demonstrating it, the SE activities where they perceive empathy as useful or less relevant, and the other factors that influence empathy of practitioners. \textcolor{black}{Rather than only observing the display or outcomes of empathy, our focus is on understanding how empathy is perceived by software practitioners. By perception of empathy, we refer to how practitioners define, interpret, and experience empathy in SE. While prior research has explored the display of empathy and its effects, our study seeks to deepen this understanding by examining what empathy means to practitioners and how it influences SE practices. This aligns with our constructivist stance as explained in Section \ref{sec:data analysis}.} By gathering insights from practitioners with diverse backgrounds, we aim to provide a nuanced understanding of empathy in SE. 

The findings of this study contribute to both research and practice. From a research perspective, the study expands the discourse on empathy in SE by capturing a wide range of practitioner perspectives. From a practical standpoint, understanding empathy in SE can inform strategies for fostering more effective teamwork, communication, and decision-making in SE environments. The key contributions of this work include the identification of:
\begin{itemize}
    \item the meaning of empathy for software practitioners both in general and within SE contexts;
    \item the motivations for demonstrating empathy in SE;
    \item the SE activities where empathy is perceived as useful or not;
    \item the other factors that influence practitioners' empathy;
    \item key future research directions for SE researchers; and
    \item practical implications for the software industry.
\end{itemize}



\section{Related Work} \label{sec:related work}

Empathy has been examined across diverse disciplines, including evolutionary, psychology, neuroscience, moral perspectives, as well as political, economic, and cultural viewpoints \cite{guthridge2020critical}. This wide-ranging interest has led to numerous definitions and conceptualisations, resulting in a lack of consensus on what empathy truly entails. Guthridge and Giummarra conducted a comprehensive analysis of 146 empathy definitions and identified six core dimensions of empathy: cognitive, affective, experience, ability, understanding, and self–other distinction \cite{guthridge2021taxonomy}. They proposed that empathy involves ``the ability to experience affective and cognitive states of another person, whilst maintaining a distinct self, in order to understand the other.'' This aligns with the widely accepted view that empathy consists of at least two primary dimensions: cognitive (e.g., ``I understand how you feel'') and affective (e.g., ``I feel what you feel'') \cite{decety2011neuroevolution, hein2008feel}. These two dimensions involve different brain regions such as the mirror neuron system for affective empathy and multiple brain regions for cognitive empathy \cite{yu2018dual}. However, they are not entirely independent but work both independently and interdependently \cite{cuff2016empathy, lumma2020insights}. Although empathy is recognised as a complex, multilayered, interconnected neuropsychological process, the underlying mechanisms that connect its cognitive and affective dimensions are still not fully understood \cite{guthridge2023role}. Notably, most of this work has focused on general or non-technical domains, leaving a gap in understanding how empathy manifests in professional settings characterised by high technical complexity and cognitive demands, such as SE.

\textcolor{black}{Previous research on empathy in SE has explored several related areas. For instance, some research has examined the role of empathy through user experience (UX) design tools such as personas and empathy maps \cite{karolita2023personas, ferreira2015designing}, while others have focused on design thinking methodologies to aid requirements elicitation \cite{canedo2020design}. In addition, some work has investigated collective empathy within team dynamics and project management contexts \cite{akgun2015collectiveempathy}. More recently, research has explored how developers demonstrate empathy in online interactions with users, using the Perception-Action Model (PAM) to classify empathic behaviours \cite{devathasan2024deciphering}. This research found that a significant proportion of online developer-user interactions exhibited elements of empathy. Another study found that empathy played a key role in helping diverse student teams navigate conflicts and improve collaboration \cite{devathasan2025empathy}.} 
\textcolor{black}{In our previous work, we developed a taxonomy for empathy models in SE \cite{gunatilake2023empathy} by analysing models from other disciplines, marking one of the first efforts to examine empathy in the SE context. To investigate empathy in real-world SE projects, we conducted a case study with a software development team, identifying enablers and barriers to empathy in developer-user interactions and strategies to overcome these barriers \cite{gunatilake2024enablers}. To make these insights more actionable for industry, we subsequently developed guidelines to help practitioners harness empathy to improve SE practices \cite{gunatilake2025guidelines}. However, these studies did not explore the broader impact of empathy in SE. To address this gap, we developed a theory capturing the multifaceted role of empathy in developer–stakeholder interactions \cite{gunatilake2025theory}. Our theory offers insights into: (a) the \textit{context} in which empathy emerges; (b) the \textit{conditions} that shape it; (c) the \textit{causes}, identifying the drivers of empathy and its absence; (d) the \textit{consequences}, outlining the outcomes of both the presence and absence of empathy; (e) the \textit{contingencies}, which include strategies to foster empathy or address barriers to it; and (f) the \textit{covariances}, representing the relationships between these categories. In this study, we identified the factors influencing empathy, which were subsequently incorporated as the \textit{conditions} of our theory. While there is some overlap with the current study in identifying the factors influencing empathy, the current study extends these findings by exploring how empathy is interpreted, the motivations behind it, and the SE activities where it is more or less relevant.}

Recent research has begun to address how empathy is conceptualised within SE. Studies have identified conceptual themes drawn from practitioner explanations such as understanding, perspective-taking, embodiment, compassion, and emotional sharing \cite{cerqueira2024empathy, cerqueira2023thematic}. These themes have been grouped into three categories: \textit{cognitive empathy} (understanding, perspective-taking, and embodiment), \textit{emotional empathy} (emotional sharing), and \textit{compassionate empathy} (compassion). 
%
However, a comprehensive understanding of its conceptualisation within SE remains limited. In particular, questions around \textit{how empathy is interpreted}, \textit{where it is considered useful or less relevant}, \textit{what motivates its expression}, and \textit{which factors influence it} have not been fully addressed. Further, much of the existing work has drawn primarily from grey literature sources, such as developer blogs. As a result, the day-to-day manifestations of empathy in professional SE settings have not been sufficiently examined from an empirical standpoint.
Our work builds on and extends this body of research by offering an empirically grounded account of empathy manifestations in SE by exploring how empathy is understood, demonstrated, and influenced in diverse SE contexts. 
This broader view contributes to a more nuanced understanding of empathy in SE and offers a foundation for future empirical studies, interventions, and measurement tools tailored to the SE context.

\section{Research Methodology} \label{sec:research design}
We employed an interview study\footnote{Approved by Monash Human Research Ethics Committee. ERM Reference Number: 41060} and a qualitative survey\footnote{Approved by Monash Human Research Ethics Committee. ERM Reference Number: 45708} to explore the manifestations of empathy in SE including how practitioners define empathy, the motivations for demonstrating empathy, SE activities where empathy is perceived useful or \textcolor{black}{less} relevant, and other factors that influence empathy of practitioners. We wanted to answer the following key research questions (RQs):
\begin{itemize}[leftmargin=2.5em]
    \item[\textit{RQ1}] \textit{What is the meaning of empathy for software practitioners, both in general and within SE contexts?:} \textcolor{black}{Differentiating empathy in SE from general contexts is crucial, as SE is a unique, highly technical field where collaboration, communication, and problem-solving dynamics differ from those in other professions. In SE, empathy not only involves emotional understanding but also supports effective collaboration, team communication, and stakeholder relationships, all within the context of time pressures, technical challenges, and complex interactions. These contextual factors shape how empathy is perceived in SE, making it distinct from its use in general contexts. By clarifying these distinctions, we can better understand how empathy functions within the unique socio-technical dynamics of SE.}
   
    \item[RQ2] What motivates software practitioners to demonstrate empathy in their work?: \textcolor{black}{This explores the underlying reasons why software practitioners choose to express empathy in their professional interactions. It focuses on identifying the motivations that drive practitioners to engage in empathetic behaviours.}

    \item [RQ3] In which specific SE activities do practitioners perceive empathy to be particularly useful or less relevant?: \textcolor{black}{Investigates which SE activities are most or least conducive to empathy, highlighting the contextual relevance of empathy across different SE practices.}

    \item[RQ4] What factors influence the empathetic behaviours of software practitioners in SE?: \textcolor{black}{Focuses on the various factors that influence empathy. 
    By examining these elements, RQ4 provides insights into how both individual characteristics and external circumstances influence the expression of empathy in SE contexts.}
    
\end{itemize}


\begin{figure*} [htbp]
    \centering
    \includegraphics[width=\textwidth]{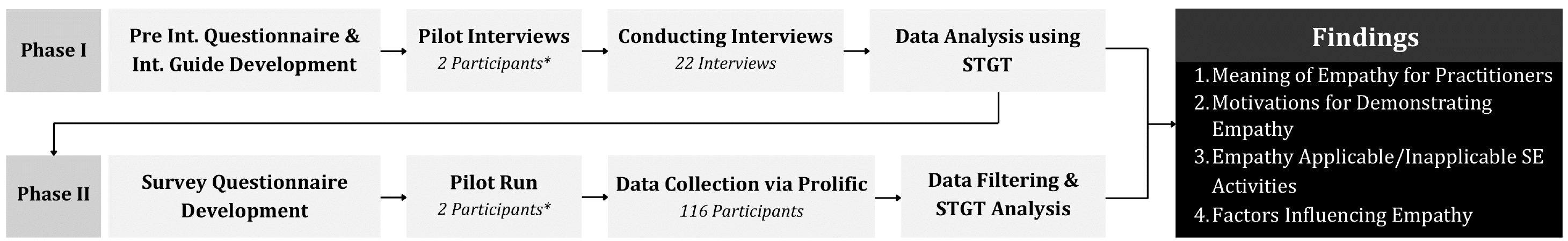}
    \caption{Overview of Research Methodology: Data collection in two phases and applying Socio-Technical Grounded Theory \cite{hoda2024qualitative} for qualitative data analysis (*Pilot study participants were not included in final analysis; Int = Interview).}
    \label{fig:methodology}
\end{figure*}

\begin{table*}[htbp]
    \caption{Overview of Participant Demographics and Team Contexts (Interviews and Surveys)}
    \label{tab:Demographic Information of the Practitioners}
    \resizebox{\textwidth}{!}{
    \begin{tabular}{@{}llllll@{}}
        \toprule
        \textbf{Role} & \textbf{\# of Practitioners} & \textbf{Age} & \textbf{\# of Practitioners} &  \textbf{Experience} & \textbf{\# of Practitioners} \\
        \midrule
        
        Developer & \blackbar{26.5}{53} & 20-30 & \blackbar{20}{40} & Less than 1 year & \blackbar{1.5}{3}\\
        
        Stakeholder & \blackbar{44.5}{85} & 31-40 & \blackbar{22.5}{55} & 1-2 years & \blackbar{2}{4} \\

        & & 41-50 & \blackbar{12.5}{25} & 3-5 years & \blackbar{22}{44}\\
        
        \cmidrule(r){1-2} 
        
        \textbf{Gender} & \textbf{\# of Practitioners} & 51-60 & \blackbar{5.5}{11} & 5-10 years & \blackbar{23}{46}\\

        \cmidrule(r){1-2} 
        
        Woman & \blackbar{24.5}{49} & 61-70 & \blackbar{2}{4} & 10-15 years & \blackbar{9}{18}\\
        Man & \blackbar{44.5}{89} & Above 70 & \blackbar{1}{2} & 15-20 years & \blackbar{4.5}{9} \\
        & & Prefer not to say & \blackbar{0.5}{1} & 20-30 years & \blackbar{4.5}{9}\\
        & & & & 30-40 years & \blackbar{2.5}{5}\\
        
        \cmidrule(r){1-6} 

        \textbf{Team Size} & \textbf{\# of Practitioners} & \textbf{Org. Size} & \textbf{\# of Practitioners} & \textbf{Development Method Used} & \textbf{\# of Practitioners}\\
        
        \cmidrule(r){1-6} 

         Less than or equal to 5 & \blackbar{15.5}{31} & Startup & \blackbar{3}{6} & Agile - Scrum & \blackbar{49.5}{99}\\

         5 - 10 & \blackbar{30}{60} & Small & \blackbar{13.5}{27} & Agile - Kanban & \blackbar{22.5}{45}\\

         10 -20 & \blackbar{16.5}{33} & Medium & \blackbar{19.5}{39} & Traditional (Waterfall) & \blackbar{35}{70}\\

         More than 20 & \blackbar{7}{14} & Large & \blackbar{33}{66} & XP & \blackbar{11.5}{23}\\

         \cmidrule(r){1-6} 
         \textbf{Country} & \textbf{\# of Practitioners} & \textbf{Country} & \textbf{\# of Practitioners} & \textbf{Country} & \textbf{\# of Practitioners}  \\
         \cmidrule(r){1-6} 

         USA & \blackbar{19.5}{39} & Germany & \blackbar{2.5}{5} & Ireland & \blackbar{1}{2} \\
         UK & \blackbar{9.5}{19} & Mexico & \blackbar{2}{4} & Sri Lanka & \blackbar{1}{2}  \\
         Canada & \blackbar{7}{14} & Chile & \blackbar{1.5}{3} & Austria & \blackbar{0.5}{1}  \\
         Australia & \blackbar{5.5}{11} & France & \blackbar{1.5}{3} & Netherlands & \blackbar{0.5}{1}  \\
         Italy & \blackbar{4}{8} & Greece & \blackbar{1.5}{3} & Poland & \blackbar{0.5}{1}  \\
         New Zealand & \blackbar{3.5}{7} & India & \blackbar{1.5}{3} & Sweden & \blackbar{0.5}{1}  \\
         Portugal & \blackbar{3.5}{7} & Hungary & \blackbar{1}{2} & &  \\
            
        \bottomrule
    \end{tabular}
    }
    \begin{flushleft}
        \textit{Additional details regarding the interview participants can be found in our previous study \cite{gunatilake2025theory}.}
    \end{flushleft}
\end{table*}

\subsection{Data Collection} \label{sec:data collection} 
We employed a multi-method qualitative approach, combining semi-structured interviews and a qualitative survey for data collection. 
Figure \ref{fig:methodology} shows an overview of our study methodology.\footnote{Indonesian participants were excluded due to ethical requirements. Monash University Human Research Ethics Committee (MUHREC) advised that foreign researchers conducting research in Indonesia must obtain prior approval from both MUHREC and the National Research and Innovation Agency (BRIN), which complicated the recruitment process.} \textcolor{black}{First, we conducted 22 interviews with software practitioners \cite{gunatilake2025theory} and used a pre-interview questionnaire to collect demographic information from the participants. We developed the interview guide by following the guidance and tips on drafting interview questions outlined in \cite{hoda2024qualitative}.} The study began with convenience sampling, followed by theoretical sampling as the study progressed. Participants were recruited through professional networks such as LinkedIn and X (formerly Twitter), as well as through personal connections. We also employed a snowballing approach, encouraging participants to refer colleagues or friends to the study. \textcolor{black}{We intentionally targeted individuals from diverse backgrounds, including developers at various experience levels and stakeholders in different roles, spanning a wide range of geographical locations. Given the complex and multifaceted nature of empathy, we believed that including participants from varied backgrounds would yield richer and more nuanced insights. This approach aligns with our constructivist research paradigm, which emphasises the importance of context and perspective in understanding complex phenomena (Section \ref{sec:data analysis}).}

\textcolor{black}{Each prospective participant's eligibility was assessed before they received the pre-interview questionnaire. For instance, since the interview focused on interactions between developers and stakeholders, participants who did not engage in such interactions were excluded from the study.} The interviews explored participants' perceptions of empathy, both in general and within SE contexts, particularly in their interactions with developers and stakeholders. In addition, we examined the factors influencing empathy in SE. During our analysis, we also identified participants' motivations for demonstrating empathy, as well as the SE activities where they considered empathy to be useful or not.

To complement the in-depth perspectives gained from the interviews and to examine whether these insights hold across a more diverse and geographically distributed sample, we subsequently conducted a large-scale qualitative survey with 116 software practitioners, recruited through Prolific\footnote{https://www.prolific.com/}. We selected Prolific as our recruitment platform due to its ability to provide a diverse and pre-screened participant pool while allowing precise control over sampling criteria. Unlike professional networks such as LinkedIn or personal contacts, which may introduce bias by limiting the sample to specific communities or professional circles, Prolific enables access to a wide range of participants across different regions, industries, and levels of experience. In addition, Prolific maintains rigorous participant screening and verification mechanisms, ensuring that respondents meet predefined qualification criteria and reducing the risk of fraudulent or low quality responses. We applied specific screening criteria on Prolific to ensure the selection of relevant participants for our study. Participants in our study were required to work in the information technology sector, have at least three years of professional experience, and regularly interact with colleagues. \textcolor{black}{We required participants to have at least three years of experience to ensure they had sufficient exposure to the complexities of SE and could provide nuanced insights into the socio-technical challenges inherent in the discipline.} To further enhance diversity in geographical representation and gender distribution, we applied iterative, purposive sampling, adjusting selection criteria across different recruitment batches. To ensure data quality, we incorporated attention-check questions to verify participant engagement and minimise the risk of random or insufficiently considered responses. In addition, all open-ended questions were made mandatory to encourage thoughtful and comprehensive answers. We excluded all responses that failed the attention-checks. Table \ref{tab:Demographic Information of the Practitioners} provides key demographic information of interview and survey participants. 

\textcolor{black}{Our 22 interviews explored both the role of empathy and its conceptualisations in SE. In our previous work, data on the role of empathy were used to develop a theory \cite{gunatilake2025theory}, while the current study focuses on the conceptualisation of empathy. The only overlap between the two studies concerns the factors influencing empathy. Of the 22 interviews, 12 provided insights into these factors, reported in the earlier study as part of theory development \cite{gunatilake2025theory}. In this study, we revisit these factors in the Section \ref{sec:Other Factors that Influence Empathy} (Other Factors Influencing Empathy), incorporating additional factors from the survey data to expand our understanding of empathy in SE. It is also important to note that a single interview can yield multiple key categories, as one interview often provides rich insights into various aspects of empathy. Thus, while there is some overlap, the current study presents new findings from both interview and survey data. Apart from the overlap, all other findings are original and not reported elsewhere.}

\begin{figure*} [htbp]
    \centering
    \includegraphics[scale=0.7]{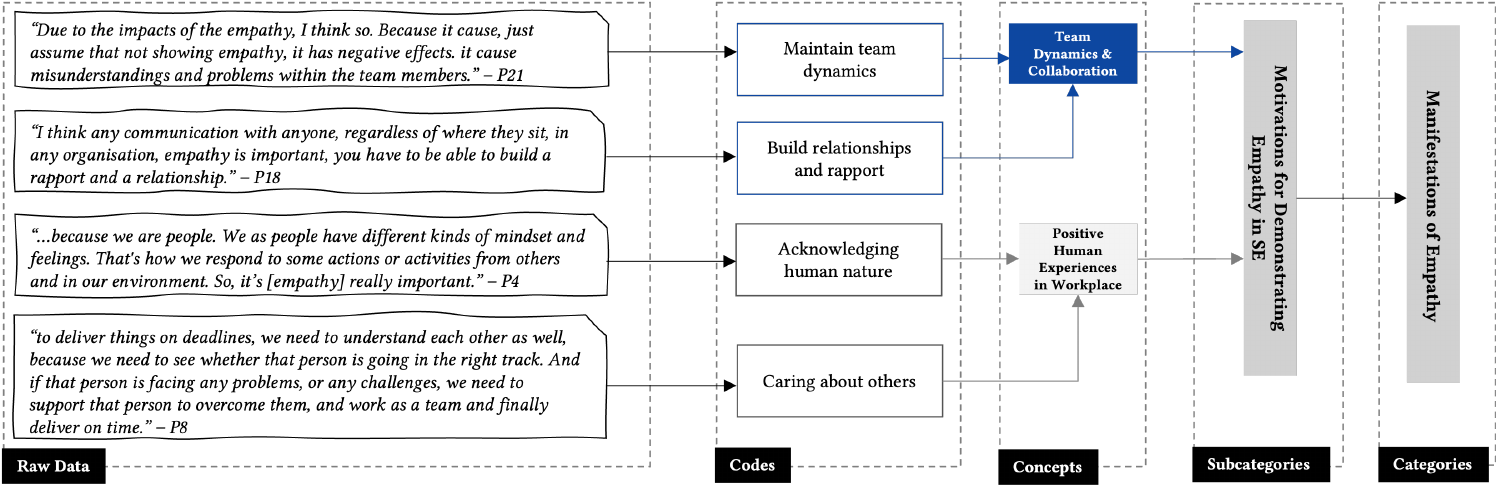}
    \caption{Emergence of the category `\textit{Manifestations of Empathy}' from raw data → codes → concepts → subcategory → category through constant comparison}
    \label{fig:stgt example}
\end{figure*}

\subsection{Data Analysis} \label{sec:data analysis}
\textcolor{black}{Socio-Technical Grounded Theory (STGT) has been formalised as a method particularly suitable for technology-intensive domain studies such as SE \cite{hoda2022STGT, hoda2024qualitative}. STGT is particularly suited for exploring socio-technical phenomena, which is why we chose it to analyse the phenomenon of empathy manifestations in SE. Further, our study aligns closely with STGT's socio-technical framework, making it an appropriate choice for our analysis. STGT consists of two stages: a basic stage for data collection and analysis, and an advanced stage for theory development. STGT allows for limited application, such as using the basic stage to analyse data from various methods. For this study, we employed STGT in its limited application to analyse the data collected from both interviews and survey study \cite{hoda2024qualitative}.} 

\textcolor{black}{STGT does not prescribe a specific research paradigm, such as positivism. Instead, it encourages researchers to select and articulate their own perspective, drawing on their expertise to interpret the data \cite{hoda2024qualitative}. The first author, who conducted the data collection and analysis, has industry experience in software development and has worked closely with both technical and business stakeholders. We acknowledge the adoption of a subjective, context-specific, constructivist research paradigm for formulating the interview questions, interpreting responses, and developing concepts, categories, and their interrelationships.}
\textcolor{black}{To strengthen the depth and breadth of our findings, we integrated qualitative data from both the interview and survey studies. The interview data provided rich, in-depth insights into how software practitioners perceive and apply empathy in their work, while the survey responses offered complementary insights from a broader and more diverse participant pool. These key concepts identified during interviews directly informed the survey design. By aligning the survey questions with these concepts, we ensured that the survey could capture the broader applicability of these findings across a larger sample.}
The data integration followed a triangulation strategy: findings from the interview study informed the survey design and later served as a reference point for interpreting the survey results \cite{storey2024guiding}. The survey data, in turn, helped to both expand upon and corroborate the patterns identified in the interviews. This combined analysis enabled us to identify consistent patterns across both datasets and to highlight areas where participants' perspectives converged or diverged. The findings are presented in an integrated manner, drawing on evidence from both interviews and surveys, to offer a more comprehensive and empirically grounded understanding of how empathy manifests in SE. This integrated approach enabled a more nuanced and robust interpretation of empathy, addressing both individual experiences and broader trends across the SE community.

We began by transcribing the interview recordings and subsequently stored and analysed the data using NVivo software. For the survey data, we first applied an initial filtering process by removing all responses that failed the attention-check questions to ensure data integrity.
Following this, we conducted a secondary round of response-level filtering, where we manually reviewed participants' answers to ensure meaningful and relevant contributions. This step was necessary to exclude low quality responses, such as those containing irrelevant, nonsensical, or overly vague answers that did not contribute meaningful insights to our analysis. This filtering was performed at two levels: at the participant level, where entire responses were removed if they consistently lacked substantive engagement; and at the question level, where specific answers were excluded while retaining the participant's valid contributions to other questions.
This additional manual verification step was essential to ensure the credibility and trustworthiness of our findings, as the initial filtering alone is insufficient to identify all low quality responses. In cases where a participant's responses were consistently vague, off-topic, or nonsensical across multiple questions, we deemed their entire submission unreliable and removed it. However, some participants provided meaningful responses to most questions but gave low quality answers to specific questions, such as one-word responses, overly generic statements, or irrelevant remarks that did not contribute meaningful insights. In these cases, we retained their valid contributions while excluding only the non-meaningful portions. By carefully distinguishing between entirely unreliable responses and those with partial value, we preserved as much high quality data as possible while ensuring that our analysis was not compromised by low quality answers. This selective data cleaning helped us to strike a balance between data integrity and inclusivity, ensuring that meaningful participant insights are retained without introducing noise into the analysis. 
As a result, we retained 36 survey responses for analysing empathy definitions, motivations, and beneficial SE activities. However, all 116 survey responses were found to be valid for examining factors influencing empathetic behaviours. In addition, we obtained valid responses on all these aspects from our 22 interviews. Consequently, our final dataset comprised 138 responses for factors influencing empathy and 58 responses for these other aspects.
Participants involved in the pilot studies were excluded from the final analysis, as these pilots were conducted to refine and improve the clarity, relevance, and structure of the interview and survey instruments. Including pilot data could have impacted the validity of the results, as they were collected under evolving conditions rather than the finalised study design.

\textcolor{black}{After data filtering, we applied the STGT for data analysis \cite{hoda2022STGT, hoda2024qualitative} to analyse our interview transcriptions and survey responses. The first author conducted open coding on all data, developing a codebook, which was peer-reviewed and refined based on feedback from co-authors, including the third author, a grounded theory expert. Following these discussions, the first author refined the codebook and conducted constant comparison. Constant comparison involved comparing codes both within individual transcripts and across different transcripts. Similar codes were grouped into concepts, related concepts were further grouped into subcategories, and similar subcategories were grouped to form categories. Memos were written throughout the process to capture the researcher's reflections, documenting emerging concepts and their interrelationships. An example of the STGT analysis is illustrated in Figure \ref{fig:stgt example}, and a sample memo in Figure \ref{fig:memo}. For a more comprehensive example of the analysis, the supplementary information package is accessible online.\footnote{https://doi.org/10.5281/zenodo.17090534}}


\begin{figure} [htbp]
    \begin{boxA}  
        \scriptsize
        Empathy in SE seems to be driven by three key motivations: enhancing team dynamics, ensuring project and business success, and fostering positive human experiences in the workplace. The most significant motivation, related to team dynamics, includes bridging technical knowledge gaps (6 IP, 4 SP), supporting self-learning (2 IP, 4 SP), and maintaining healthy relationships (6 IP, 16 SP), all of which contribute to better collaboration. Empathy is also seen by some participants as vital in project success (4 IP, 10 SP), as it helps practitioners understand user needs and tailor solutions effectively (7 IP, 3 SP), preventing issues and improving outcomes. Further, empathy is seen to contribute to a positive workplace culture by prioritising well-being (4 IP, 9 SP), acknowledging human needs (7 IP, 1 SP), and ensuring that team members' perspectives and emotions are considered (1 IP, 10 SP). 
    \end{boxA}  
    \caption{Memo on ``Motivations for demonstrating empathy in SE''}
    \label{fig:memo}
\end{figure}

    
    

\section{Findings} \label{sec:findings}
\begin{figure*} [htbp]
    \centering
    \includegraphics[width=\textwidth]{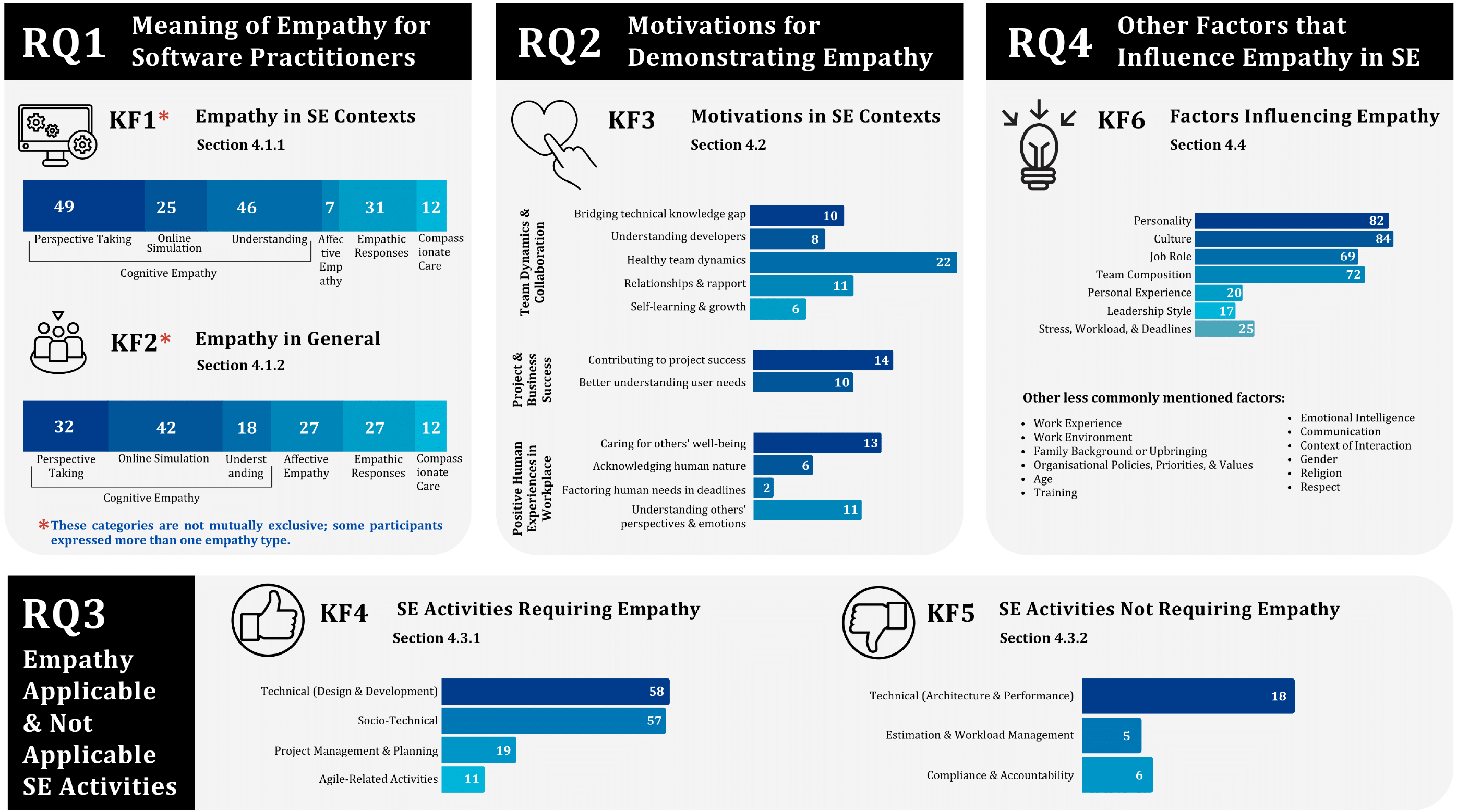}
    \caption{Overview of key findings highlighting empathy definitions, motivations, SE activities, \& influencing factors based on interview and survey data}
    \label{fig:Overview of key findings}
\end{figure*}

An overview of our key findings is illustrated in Figure \ref{fig:Overview of key findings}. In the following sections, we share several quotations from participants that serve both as evidence of the underlying raw data and as a way for the readers to experience the phenomenon first-hand from the participants' perspective. However, due to confidentiality concerns, we cannot share all underlying quotations. Given the large number of participants (n=138), we do not assign individual identifiers (e.g., P1, P2). Instead, we indicate the data source by referring to participants as IP (interview participant) or SP (survey participant), along with the number of participants who expressed similar views. 


\subsection{Meaning of empathy for practitioners (RQ1)} \label{sec:Meaning of empathy} 
\textcolor{black}{We asked participants to define what empathy means when interacting with software developers or stakeholders, as well as how they understand empathy in general.}
Our analysis revealed that participants' definitions of empathy primarily referred to four key categories: \textit{cognitive empathy}, \textit{affective empathy}, \textit{empathic responses (behavioural empathy)}, and \textit{compassionate care}. 

\subsubsection{Empathy in SE Context} \label{sec:Empathy in SE Context}
We found that participants' definitions of empathy aligned with all above key categories in the context of their interactions with developers and stakeholders in SE.

\textcolor{black}{\textbf{Cognitive empathy} is the ability of a person to consciously detect and understand the internal states of others \cite{clark2019feel}. Cognitive empathy in SE contexts emerged in three distinct forms: \textit{perspective taking} defined as the ability to cognitively understand and adopt another person's perspective, considering their thoughts, beliefs, and experiences, without necessarily engaging with their emotions \cite{reniers2011QCAE}; \textit{online simulation} defined as the ability to mentally represent and understand how another person might feel in a given situation, without necessarily experiencing the emotion oneself. This process allows individuals to cognitively simulate emotional states, such as stakeholders imagining user needs, without directly engaging in the emotional experience of those needs \cite{reniers2011QCAE}; and \textit{understanding} defined as the ability to comprehend and interpret emotions, intentions, behaviours, and experiences of others.} 


\textit{Perspective taking} emerged as the most closely related type of empathy for software practitioners, cited by all 22 IP and 27 SP (n=49). Participants emphasised its significance across various interactions, including understanding stakeholder needs, motivations, and business goals, collaborating with others (e.g., testers, project managers), recognising developers' challenges, workload, and priorities, empathising with end-user frustrations, and evaluating usability from user's perspective. In addition, they highlighted the importance of understanding how others' ideas fit into the broader context, valuing project timelines, and adapting to different developers' working styles and technical approaches.

\begin{quote}
    \small
    \faIcon{comments} \textcolor{black}{[IP]} \textit{``in the context of work, empathy is being able to understand their problems or why they might want to solve a particular problem. Just being able to understand or to figure out, like if they feel a certain way, and why that might be.''}
    
\end{quote}

\textit{Online simulation} emerged as a key aspect of empathy in SE, identified by 9 IP and 16 SP (n=25). Participants described this form of empathy in various contexts, including stakeholders imagining themselves in users' positions to understand their needs, mentally simulating the emotional and psychological states of project managers, imagining others' experiences to respond appropriately, considering the impact of work-life balance on colleagues, and stepping into others' shoes to better understand their perspectives and challenges.

\begin{quote}
    \small
    \faIcon{comments} \textcolor{black}{[IP]} \textit{``To me, empathy is trying to step into the shoes of the different people that I work with.''}
\end{quote}

Thirteen IP and 23 SP (n=36) provided definitions aligned with \textit{understanding}. They emphasised the importance of grasping stakeholder needs before making judgments, understanding different ideas and job roles, and recognising how stakeholders' contributions fit into the broader context. Participants also highlighted the need to understand the rationale behind specific features, user and stakeholder needs, developers' thought processes, working patterns, development dependencies, and the significance of tasks or issues for their colleagues.
\begin{quote}
    \small
    \faIcon{comments} \textcolor{black}{[IP]} \textit{``trying to understand that it's their job to sort of take away the problems that users experience rather than treating things like a technical problem that can be overlooked.''}
\end{quote}

\textbf{Affective empathy} is \textcolor{black}{the ability} of a person to perceive and share other individual's emotional states and feelings \cite{clark2019feel, reniers2011QCAE}. Affective empathy emerged in participants' responses as the ability to experience self-oriented emotions in response to others' emotional states. Only a small number of participants, including 6 IP and 1 SP (n=7), referred to affective empathy, describing their experience of absorbing positive and negative emotions and deeply feeling why others experience certain emotions.
\begin{quote}
    \small
    \faIcon{comments} \textcolor{black}{[SP]} \textit{``Empathy for me in this is understanding that not everyone is like me and i must be able to see and feel what others are going through...''}
\end{quote}

\textbf{Empathic responses} or behavioural empathy is the outward expression of empathy through actions and behaviours \cite{clark2019feel}. It was cited by 10 IP and 21 SP (n=31). Participants described various empathy-driven actions and behaviours, including thoughtfully responding to others' needs and concerns, fostering collaboration, and acknowledging the complexity of developers' technical work or the pressure they face under tight deadlines. Other empathic behaviours included aligning with colleagues' needs rather than pushing personal agendas, actively listening without judgment, and tailoring solutions to both practitioners' and business needs. In addition, participants emphasised the importance of effectively communicating technical information, adjusting language based on context, and creating a positive and productive work environment. Empathy was also reflected in thoughtful task reviews, valuing practitioners' knowledge and skills, avoiding unnecessary criticism or micromanagement, respecting others' time, providing clear requirements, and dedicating time to support colleagues.
\begin{quote}
    \small
    \faIcon{comments} \textcolor{black}{[SP]} \textit{``Not jumping to conclusions when they have issues in the development process and debugging. Also it means setting aside time to help others blow off steam because development jobs can be stressful and irritating, especially when debugging or dealing with incomplete requirements and colleagues who don't have an understanding of the development process or are generally toxic.''} 
\end{quote}

\textbf{Compassionate care} is the tendency of a person to experience feelings of warmth, compassion and concern for others undergoing negative experiences \cite{clark2019feel}. It emerged in the responses of 10 IP and 2 SP (n=12). Participants described it as actively caring for colleagues' well-being, being attentive to others' emotions, and demonstrating genuine human interest in their experiences. They also highlighted the importance of offering support during personal or professional challenges, as well as considering end-users' experiences and being mindful of their emotions.
\begin{quote}
    \small
    \faIcon{comments} \textcolor{black}{[IP]} \textit{``I just show my empathy, try my best to let them know, someone care about you, is not like  stay in a corner and just face a screen to type in the program all day.''}
\end{quote}

Our findings reveal a \textit{strong similarity in how developers and stakeholders perceive and define empathy} within the SE context. \textcolor{black}{The majority of developers (n=37) and stakeholders (n=56) primarily associated empathy with cognitive empathy. While this was the most common framing, some participants also referred to affective empathy. Specifically, one developer and six stakeholders additionally mentioned affective empathy.}
This suggests that software practitioners, whose roles are largely technical and task-oriented, may naturally adopt a more analytical approach to empathy, focusing on problem-solving and understanding requirements over emotional engagement. Empathic responses emerged as the second most frequently mentioned category, identified by 14 developers and 17 stakeholders, highlighting that while practitioners primarily engage with empathy cognitively, they also recognise the importance of translating understanding into concrete actions, such as collaboration, clear communication, and thoughtful support. Compassionate care was the third most prevalent category, referenced by three developers and nine stakeholders, indicating that although emotional engagement is less central in SE, practitioners still acknowledge the value of fostering a supportive work environment and considering the well-being of both colleagues and end-users.

\subsubsection{Empathy outside SE Contexts} \label{sec:Empathy in General}
Compared to software practitioners' definitions of empathy within SE contexts, we observed a notable shift in emphasis when considering empathy in other settings, particularly with a greater focus on \textit{affective empathy}. While only seven participants mentioned affective empathy in SE contexts, 27 participants referred to it in broader, non-SE contexts. In addition, we found a significant increase in the number of participants whose definitions aligned with \textit{online simulation} (n=42). Conversely, there was a notable decline in participants referencing \textit{perspective taking} (n=32) and \textit{understanding} (n=18), as well as a slight decrease in mentions of \textit{empathic responses} (n=27). The number of participants referencing \textit{compassionate care} remained unchanged (n=12). These shifts suggest that while empathy in SE is predominantly framed in cognitive and action-oriented terms, outside of SE contexts, practitioners place greater emphasis on emotional connection and the ability to mentally simulate or deeply relate to others' experiences.

For \textit{perspective taking}, participants described looking at situations or problems from someone else's point of view, understanding the reasons behind others' requests, recognising their circumstances, taking the time to consider different viewpoints, acknowledging difficulties, and respecting differing opinions.
For \textit{online simulation}, they provided examples of putting themselves in others' shoes, understanding what others are going through, grasping their emotions, and imagining the paths others take in various situations.
For \textit{understanding}, participants mentioned examples of grasping others' reactions, feelings, struggles, experiences, and the situations they are dealing with.
%
Regarding \textit{affective empathy}, participants shared experiences of understanding and feeling others' emotional states, empathising with their pain, sharing in their emotions, and reacting emotionally to the circumstances others were experiencing.
In terms of \textit{empathic responses}, participants highlighted actions such as fostering connections, engaging in meaningful communication, actively listening without judgment, responding thoughtfully to others' situations, offering support, and helping resolve others' problems.
In the context of \textit{compassionate care}, participants spoke about caring for others as human beings, attending to their personal emergencies, recognising others' limitations, and offering emotional support during challenging times.

\begin{quote}
    \small
    \faIcon{comments} \textcolor{black}{[IP] \textit{``..empathy is all about being able to step into someone else's shoes, or to look at a situation or a problem or a question or whatever, from someone else's perspective, and being able to understand whatever reaction that they might have, to that given situation, a question or a problem.''}}
    
    \faIcon{comments} \textcolor{black}{[SP] \textit{``When I hear the word empathy, the first thing that comes to mind is stepping into someone else's shoes, not just understanding their situation but truly feeling what they might be going through. It's that moment when you pause your own judgments, set aside your own experiences, and connect with someone on a deeper emotional level.''}}
\end{quote}



\subsection{Motivations for Practising Empathy (RQ2)} \label{sec:Motivations} 
Our analysis revealed the underlying motivations that drive practitioners to demonstrate empathy, underscoring its perceived importance and value in SE contexts. These motivations were organised into three key subcategories: \textit{team dynamics and collaboration}, \textit{project and business success}, and \textit{fostering positive human experiences in the workplace}.
The most commonly cited motivations fell under \textit{team dynamics and collaboration}, including bridging the technical knowledge gap (6 IP, 4 SP), understanding developers (5 IP, 3 SP), maintaining healthy team dynamics (6 IP, 16 SP), building relationships and rapport (3 IP, 8 SP), and supporting self-learning and growth (2 IP, 4 SP).
\begin{quote}
    \small
    \faIcon{comments} \textcolor{black}{[SP]} \textit{``I'm motivated to demonstrate empathy because it helps build stronger relationships, improves communication, and leads to better collaboration.''}
\end{quote}

Motivations associated with \textit{project and business success} included contributing to overall project success (4 IP, 10 SP) and better understanding user needs (7 IP, 3 SP).
\begin{quote}
    \small
    \faIcon{comments} \textcolor{black}{[SP]} \textit{``Having a level of empathy for users when a problem arises allows me to tailor my line of questioning to gather necessary information, I am motivated to ask the right questions and guide people to retrieve details I need by my goal of solving issues and preventing them from reoccurring [...] Empathy factors in with the realisation that we all make mistakes in software, and that we don't have time to be mean or argue about any of it because it needs to be fixed by anyone who can do the job.''}
\end{quote}

The final category, \textit{fostering positive human experiences in the workplace}, reflects motivations that extend beyond direct project outcomes. Practitioners expressed that empathy was important for creating humane, supportive work environments. This included caring for others' well-being (4 IP, 9 SP), acknowledging human nature (5 IP, 1 SP), factoring in the human needs during deadline setting (2 IP), and understanding others' perspectives and emotions (1 IP, 10 SP).
\begin{quote}
    \small
    \faIcon{comments} \textcolor{black}{[SP]} \textit{``What motivates me to demonstrate empathy as a software practitioner is the simple fact that software development is deeply human at its core. Behind every line of code, every requirement, and every deployment are people who want their needs understood, their efforts appreciated, and their challenges acknowledged. Empathy helps bridge the gap between technical work and human connection, which leads to better outcomes for everyone involved.''}

\end{quote}

Overall, these categories illustrate that practitioners view empathy not only as a tool for individual relationships but also as a facilitator of effective teamwork, organisational success, and a more humane workplace culture.

\subsection{Empathy and SE Activities (RQ3)} \label{sec:Empathy Applied SE Activities} 

\subsubsection{SE Activities Where Empathy is Beneficial}

Our analysis of participant responses revealed several SE activities where empathy was applied and was seen to help. We categorised them into four main groups: design and development (technical), socio-technical, project management and planning, and agile-related activities. 
In \textit{design and development}, empathy played a role in tasks such as requirements discussion (6 IP, 9 SP), application design (3 IP, 5 SP), application development (11 IP, 1 SP), defect fixing (9 IP, 5 SP), testing (4 IP), code reviews and feedback (7 SP), feature handover processes (3 IP). In these activities, empathy enabled developers to align solutions with user needs and stakeholder expectations, enhancing product quality. 
In \textit{socio-technical activities}, empathy was applied during collaboration with others (9 IP, 17 SP), customer/user support and meetings (7 IP, 7 SP), team discussions (6 IP), guiding juniors (2 IP, 1 SP), understanding stakeholder needs, goals, limitations (5 SP), and understanding workloads, capacities, limitations of team members (3 SP). For these activities, empathy facilitated stronger teamwork, effective client meetings, and supportive environments, promoting open communication and trust.
In \textit{project management and planning}, empathy influenced activities such as setting project deadlines (7 IP, 1 SP), making release-related decisions (3 IP, 1 SP), managing workloads (3 IP), clarifying project direction and transitions (1 IP, 1 SP), and task prioritisation (2 IP). In these activities, empathy led to realistic deadlines, balanced workloads, and smoother project transitions.
In \textit{agile-related activities}, empathy was key during sprint planning meetings (2 IP), stand-up meetings (4 IP), and sprint retrospectives (5 IP). In these activities, empathy encouraged open dialogue during agile ceremonies, improving team alignment and continuous growth.
Overall, applying empathy in these areas was seen to foster collaboration, understanding, and user-focused outcomes.  

\begin{quote}
    \small
    \faIcon{comments} \textcolor{black}{[SP] \textit{``I think I have to show empathy when the stakeholder asks for specific implementations and deadlines. I always try my best to put myself on their shoes and understand exactly what they want and when. I also think we must be able to think like an end user above all, we are creating for users and everything has to be user friendly and intuitive.''}}
    
    \faIcon{comments} \textcolor{black}{[SP] \textit{``When something has gone wrong within a technical team, empathy is needed to avoid finger-pointing and condemnation so that the issue can be reported and not hidden out of embarrassment, and so that everyone stays on task to solve it.''}}
\end{quote}

\subsubsection{SE Activities Where Empathy is Unnecessary} \label{sec:SE Activities Not Requiring Empathy} 
We identified several SE activities where empathy was considered unnecessary, which we categorised into three areas: technical tasks, estimation and workload management, and compliance \& accountability.
In \textit{technical tasks}, such as code architecture and performance (10 SP), troubleshooting (1 IP, 3 SP), and automation (4 SP), empathy was seen as irrelevant, as these activities focused primarily on logical problem-solving. For technical tasks such as algorithm design, coding, refactoring, and optimisation, the focus was on logically identifying issues, without the need to consider others' emotions. 
In \textit{estimation and workload management}, empathy was deemed unnecessary when dealing with repeated overwork (2 IP), and unrealistic or critical deadlines (2 IP, 1 SP), where empathetic responses could hinder effective problem resolution. Here, empathy was seen as irrelevant in unreasonable conditions, as empathy in these situations could lead to negative outcomes rather than offering effective support.
In \textit{compliance and accountability activities}, empathy was not required for enforcing accountability (3 IP), and adhering to contractual obligations and procedures (3 IP) as these tasks relied on rules and policies rather than emotional considerations. In these activities, rules and policies governed actions, leaving little room for empathetic involvement.
Overall empathy was considered unnecessary in activities that typically involved objective tasks, procedural adherence, or situations where empathetic considerations do not influence the outcome.    

\begin{quote}
    \small
    \faIcon{comments} \textcolor{black}{[SP]} \textit{``..empathy is less applicable in purely technical activities like code implementation \& performance optimisation. In these tasks, focus is primarily on problem-solving, system architecture, and efficiency, where emotional understanding does not directly impact the outcome.''}

    \faIcon{comments} \textcolor{black}{[IP]} \textit{``..but in very tough deadlines and when there are like unrealistic expectations being said that you know, the next Friday is our sprint closure, we'll have to deliver this, complete things and so on. Beyond a limit, you know that this is not possible. So at some point, you tend to resist and that this cannot be done, I understand your situation, but there is a limit to it. And I will not be able to, empathise with you and do all this.''}
\end{quote}

\subsection{Other Factors that Influence Empathy (RQ4)} \label{sec:Other Factors that Influence Empathy}
Many participants highlighted various factors that influence empathy, including personality (5 IP, 77 SP), culture (5 IP, 79 SP), job role of the practitioners (3 IP, 66 SP), team composition (1 IP, 71 SP), personal experiences (20 SP), leadership style (17 SP), and stress, workload, and deadlines (25 SP). Other factors mentioned by some participants were work experience (8 SP), work environment (7 SP), and family background or upbringing (7 SP). In addition, a few participants identified factors that were less commonly mentioned, such as organisational policies, priorities, and values (6 SP), age (5 SP), training (5 SP), emotional intelligence (4 SP), communication (4 SP), and virtual vs. in-person context of interaction (3 SP). Other less-reported influences included gender (2 SP), religion (2 SP), and respect (2 SP).

Participants emphasised that \textbf{personality} plays a significant role in shaping empathy within software development. Introverted developers or those focused on technical tasks often exhibited less empathy, prioritising work over interpersonal interactions, which created barriers to engaging empathetically with stakeholders and colleagues. Conversely, those with naturally empathetic dispositions were seen to foster better team dynamics, enhance collaboration, and build trust. They shared that traits like openness, agreeableness, and conscientiousness can enhance empathetic interactions. A recurring notion was that empathy is not just a skill to be learned but is deeply rooted in personality traits. 
\textbf{Cultural context} significantly influenced empathy levels. Participants who worked in different countries observed distinct variations in how empathy was expressed and encouraged. In some cultures, employees were hesitant to express discomfort or dissatisfaction openly, while in others, such issues were addressed directly during retrospectives. Further, open and inclusive organisational culture fostered empathy, while rigid and hierarchical structures hindered it.

\textbf{Job role} also affected empathy dynamics. Participants observed that empathy varied across roles such as product owners, project managers, and testers. Tensions frequently arose between developers and project managers due to conflicting priorities, while testers struggled to maintain empathy with developers during defect identification. Developers with a narrow focus on coding often lacked empathy towards broader team needs, hindering collaboration. Roles that involved close collaboration with users typically fostered a deeper empathy for user needs, compared to more isolated roles. Such isolated roles, which limited interaction with people outside of immediate coworkers or supervisors, also contributed to reduced empathy. Empathy within job roles is shaped by shared or conflicting goals, as understanding tends to increase when roles align but decrease during disagreements.
\textcolor{black}{\textbf{Team composition} also emerged as a relevant factor in how empathy was experienced. Several participants perceived that teams with gender diversity, especially those including female members, tended to foster more open communication and frequent team interactions, which in turn encouraged empathetic engagement. In contrast, participants described some all-male teams, particularly those in back-end roles, as less communicative and more technically focused. However, it is important to emphasise that these observations reflect participants' perceptions and cannot be interpreted as evidence of a causal link between gender diversity and empathy. Other factors, such as the nature of the work (e.g., back-end versus front-end tasks), existing team norms, or broader organisational culture, may also contribute to these dynamics.
Diversity in terms of gender, culture, and background was generally seen to enrich team collaboration by bringing a wider range of perspectives. While this diversity created opportunities for empathy, it could also lead to interpersonal tensions, especially when communication styles or expectations differed. Hybrid teams, those blending multiple forms of diversity, were often perceived as effective, with participants noting stronger team cohesion and shared purpose.}

\begin{quote}
    \small
    \faIcon{comments} \textcolor{black}{[IP]} \textcolor{black}{\textit{``so the thing is, part of empathy, being able to have an ongoing empathic relationship is more than interactional skills. It is personality based in a lot of ways, but also requires respect, and trust. And if you don't have respect and trust, you are not going to have much empathy.''}}

    \faIcon{comments} \textcolor{black}{[IP]} \textcolor{black}{\textit{``I have worked in both [country x], in [country y]. but the difference I see is, in [country x] we do hesitate to raise if we were hurt from developers not having empathy. We keep it to ourselves. We might tell two to three of our friends, like he was very rude to me, but we won't sometimes tell it in our retrospective. But in here [country y], they are on point, like if they got hurt, or the developers were not empathetic, even he was not empathetic with other team members, in the retrospective they always point it.''}}
\end{quote}

\subsection{Other Findings}
\subsubsection{Empathy Reciprocation} \label{sec:Empathy Reciprocation}
An interesting insight from the interview study was the concept that \textit{``Empathy is a two-way street.''} Participants revealed that empathy in SE is most effective when it is mutual, with 13 IP highlighting that they were more likely to show empathy when it was reciprocated. A lack of empathy reciprocation was seen as a key factor contributing to empathy breakdowns \cite{gunatilake2025theory}. 
Empathy was most effective when developers and stakeholders understood each other's challenges. When stakeholders were supportive, developers reciprocated empathy, fostering collaboration. However, unempathetic stakeholders caused developers to withdraw, leading to disengagement and resistance. This lack of empathy negatively affected team morale and project outcomes, creating disengagement and isolation.
Conversely, mutual empathy created a positive cycle of collaboration, trust, and motivation. Participants emphasised that empathy must be both genuine and mutual to sustain these positive dynamics, as one-sided empathy led to broken relationships, reduced morale, and lower productivity. 
A positive cycle of support emerged when empathy was mutual, creating a cultural norm of collaboration and understanding. Developers felt motivated to go the extra mile, knowing their efforts were valued and would be reciprocated. However,  when not reciprocated, it led to broken relationships, reduced morale, and decreased productivity. Participants emphasised that empathy must be both genuine and mutual to sustain these positive dynamics.
\begin{quote}
    \small
    \faIcon{comments} \textcolor{black}{[IP]} \textit{``if stakeholders are empathetic, you also tend to be more empathetic. Okay, in these situations, they have been more empathetic towards me. now it's my time to show it back and give things. But the moment you start feeling that from the other side there's not enough empathy and support, then you also start believing that, this is not my game, and you as a personality if you're very empathetic, you can go to a limit. But beyond that, you also give up and whole product starts breaking where people move out.''}
\end{quote}

\subsubsection{Nature vs Nurture} \label{sec:empathy training}
Empathy training is a widely debated topic across various disciplines. Some researchers suggest that empathy is influenced by birth \cite{butters2010meta, warrier2018genome, abramson2020genetic}, while numerous studies support the notion of empathy training across different settings, with the majority reporting it as an effective method for promoting prosocial behaviour \cite{teding2016efficacy, butters2010meta, winter2022experiences}. In the healthcare domain, empathy training has been extensively researched \cite{dexter2012research, brunero2010review, batt2013teaching, stepien2006educating, baker1989integrating, baker1990systematic, winter2022experiences}. Beyond healthcare, empathy education and training has also been explored in areas such as engineering \cite{strobel2013empathy, hess2016voices, walther2017model}, reducing prejudice among children \cite{beelmann2014preventing}, improving child abuse prevention programs \cite{wiehe1997approaching}, and enhancing rape prevention programs \cite{foubert2006effects, o2003rape, lee1987rape}. Researchers have developed various models and techniques to help educators incorporate empathy into classroom settings \cite{cecilia2023model, haag2019exploring}, as well as for fostering empathy in professional work environments \cite{gunatilake2023empathy}.
In line with this literature, our interview participants (IP) expressed varying perspectives. One IP explicitly described empathy as an innate quality, but even they recommended integrating module on empathy and interactive skills into SE education, indirectly supporting the idea of empathy training. Subtle references to empathy as an intrinsic trait were also made by other IP (n=4). 
In contrast, 2 IP directly emphasised the importance of empathy training, while several others indirectly supported it (n=7). They noted that while students receive extensive technical training, they often lack the ability to empathise with others' perspectives. This gap can lead to ineffective problem-solving when addressing ill-defined issues. Their views align with existing literature, which highlights the need for empathy training to bridge the gap between technical expertise and interpersonal understanding in professional settings \cite{blanco2017deconstructing, cross2004expertise}.
Participants acknowledged both innate (nature) and learned (nurture) components in the development of empathy, with most attributing greater influence to learned experiences. While it is unclear which is more influential, they consistently endorsed the value of empathy training, suggesting that such initiatives would likely be well-received within the software industry.
\begin{quote}
    \small
    \faIcon{comments} \textcolor{black}{[IP]} \textit{``I think it's part of his nature as well. Because you know, empathy is not really something you can learn.''}\\
    \faIcon{comments} \textcolor{black}{[IP]} \textit{``..everybody working in IT doesn't matter what they're doing, doing IT degree also, does a module on empathy and working with customer, not against customer.''}
\end{quote}

\section{Discussion} \label{sec:discussion}
We discuss the broader implications of our findings for both research and practice in SE contexts, examining their significance and potential impact within the field.

\subsection{Implications for Research} \label{sec:Implications for Research}
\noindent \faIcon{graduation-cap} \textcolor{black}{\textbf{Developing an SE Empathy scale:} Several disciplines have developed domain-specific empathy scales to ensure greater relevance, accessibility, and interpretability for professionals within those contexts. For example, the Jefferson Scale of Physician Empathy (JSPE) was created for healthcare professionals \cite{hojat2001jefferson, hojat2016jefferson}, while the Empathy in Design (EMPA-D) scale was designed for the service design domain \cite{drouet2024development}. In our previous work \cite{gunatilake2023empathy}, we argued that these domain-specific instruments, as well as widely used generic empathy scales such as the IRI \cite{davis1980IRI, davis1983measuring} and QCAE \cite{reniers2011QCAE}, are not directly transferable to SE. This is due to fundamental contextual differences, including the nature of professional roles, communication practices, and socio-technical collaboration in SE. However, developing an SE-specific empathy scale requires a robust understanding of how empathy manifests within this domain. The current study provides this necessary foundation (Section \ref{sec:Empathy in SE Context}). Building on these insights, a tailored empathy scale for SE could enable more accurate and meaningful assessments of empathy among software practitioners and between software practitioners and users. Such a scale would not only support future empirical investigations but also facilitate the evaluation of empathy-related interventions and training in SE.}


\noindent \faIcon{graduation-cap} \textbf{Cross-cultural perspectives:} Empathy research in other disciplines has consistently shown that culture significantly influences empathic behaviour and expression \cite{jami2023interaction}. Our study similarly found that culture plays a key role in how empathy is perceived and enacted in SE (Section \ref{sec:Other Factors that Influence Empathy}). Given that software teams are increasingly global and often span multiple cultures and time zones, understanding the cultural dimensions of empathy is essential for effective collaboration. While other disciplines have established strong links between culture and empathy, there is limited research exploring this relationship in SE. \textcolor{black}{For example, a very direct style of code reviews, which may be valued in some cultures as efficient and unambiguous, could be perceived as overly critical or discouraging in other cultures. Balancing technical precision with cultural sensitivity allows feedback to remain constructive while maintaining empathy and respect for other perspectives. Similarly, while some cultures expect open discussions during retrospectives, others may be less comfortable voicing concerns in group settings, perceiving such critique as confrontational.} 
Investigating how such cultural norms shape the ways team members interpret, express, and respond to empathy could inform strategies to promote empathy in multicultural teams. Such insights may also support the development of culturally sensitive interventions that enhance communication and team dynamics in diverse software development environments.


\noindent \faIcon{graduation-cap} \textcolor{black}{\textbf{Intervention and training programs:} Researchers in other disciplines have extensively examined the effectiveness of empathy training in various settings, with most reporting that empathy training successfully promotes prosocial behaviour and increases empathy levels among professionals \cite{teding2016efficacy, butters2010meta, dexter2012research, brunero2010review}. Our findings similarly suggest that empathy training can enhance empathetic behaviours among software practitioners (see Section \ref{sec:empathy training}). However, research on empathy training within SE remains limited \cite{gunatilake2023empathy}. A recent study employed a video-based training approach and found that such training improved SE students' conceptual understanding of empathy \cite{mitrovic2025video}. Building on these early efforts, further research is needed to design and evaluate targeted interventions that cultivate empathy in SE contexts. These interventions could help practitioners improve empathy in task-specific situations where it is most relevant (see Section \ref{sec:Empathy Applied SE Activities}). By systematically testing and refining empathy training methods, researchers can identify approaches that effectively support communication, collaboration, and team dynamics. Additionally, investigating how different training formats affect practitioners based on their roles and personality traits may lead to more tailored and impactful training programs that meet the diverse needs of software teams.}

\noindent \faIcon{graduation-cap} \textcolor{black}{\textbf{Understanding the long-term dynamics of empathy in SE:}
Our study highlights that empathy positively influences team communication, collaboration, and practitioner well-being in the short term (Sections \ref{sec:Empathy Applied SE Activities}). These findings align with previous studies reporting short-term benefits of empathy in software teams \cite{gunatilake2025theory, cerqueira2024empathy}. However, current research, including our own, is largely based on cross-sectional data, offering only a snapshot of empathy in practice. This limits our understanding of how empathy develops, fluctuates, or sustains over time in real-world SE settings. The implication is that longitudinal studies are essential for advancing empathy research in SE. They could capture how empathy evolves across project lifecycles, how it is maintained (or eroded) in distributed or high-pressure environments, and how sustained empathic practice impacts long-term outcomes such as retention, resilience, and team cohesion. Such research could also inform the development of organisational practices and interventions that embed empathy into long-term team culture and SE processes, rather than treating it as a one-off training outcome.}


\subsection{Implications for Practice} \label{sec:Implications for Practice}
\noindent \faIcon{laptop} \textbf{Embedding empathy in SE processes:} \textcolor{black}{With the understanding of SE processes where empathy is useful (see Section \ref{sec:Empathy Applied SE Activities}), organisations can integrate empathy into relevant activities such as agile methodologies, code reviews, and stakeholder communication to improve collaboration and user satisfaction. By fostering empathy within these processes, teams can ensure that the perspectives of both users and fellow team members are considered \cite{gunatilake2025theory}. For example, during agile sprints, empathy can guide better understanding of user needs, leading to more user-centric product development. Similarly, in code reviews, empathy may allow for constructive feedback that fosters professional growth and mutual respect, rather than creating a competitive or adversarial environment \cite{gunatilake2025theory}. In stakeholder communication, empathy may help bridge the gap between technical and non-technical stakeholders, promoting clearer communication and a shared sense of purpose, which may improve the overall success of software projects \cite{gunatilake2025theory}}.

\noindent \faIcon{laptop} \textbf{Fostering empathy through motivational alignment:} Organisations can actively foster empathy in SE teams by recognising and leveraging the various motivations that drive practitioners to demonstrate empathy. Motivations such as enhancing team dynamics and collaboration, contributing to project and business success, and fostering positive human experiences can serve as powerful catalysts for integrating empathy into daily practices (see Section \ref{sec:Motivations}). \textcolor{black}{Specifically, fostering an atmosphere where empathy supports both personal growth and project outcomes can enhance collaboration and team cohesion \cite{gunatilake2025theory}. Through targeted initiatives and a focus on empathy as a core value, organisations can help motivate practitioners to engage in empathetic behaviours that may benefit team dynamics and the broader workplace culture.}

\noindent \faIcon{laptop} \textbf{Leadership and team dynamics:} Encouraging empathy-driven leadership and fostering psychological safety within teams can enhance trust, motivation, and overall work culture. Empathy in leadership enables leaders to understand and respond to the emotional needs of their team members, which can create a more supportive and collaborative work environment (see Section \ref{sec:Other Factors that Influence Empathy}). When leaders demonstrate empathy, they are more likely to build strong relationships with their team members, leading to greater trust and openness \cite{gunatilake2025theory}. In addition, empathy-driven leadership can help mitigate stress and burnout by offering appropriate support and recognition. Psychological safety, which is closely tied to empathy, allows team members to express themselves without fear of judgment or retribution, fostering an environment of innovation and collaboration. \textcolor{black}{This may help boost morale and motivation, leading to higher levels of job satisfaction and team performance \cite{gunatilake2025theory}.}

\noindent \faIcon{laptop} \textbf{Remote and hybrid work challenges:} We identified context of interaction (virtual vs in-person) as a factor influencing empathy (see Section \ref{sec:Other Factors that Influence Empathy}). \textcolor{black}{With the increasing shift toward remote and hybrid work environments, organisations can implement strategies to maintain empathetic connections between team members who may be geographically dispersed. This can include establishing asynchronous communication norms that allow for thoughtful responses and respect for each other's time zones, as well as structured feedback mechanisms to ensure that team members feel heard and valued. Virtual team-building activities, such as online collaborative exercises or virtual coffee breaks, can also help create a sense of camaraderie and emotional connection, even in the absence of physical presence. These strategies can help maintain empathy in remote or hybrid teams, where the lack of face-to-face interaction may sometimes lead to misunderstandings or feelings of isolation. By intentionally fostering empathy through these approaches, organisations may help bridge the emotional distance between remote workers and maintain a strong sense of team cohesion.}

\noindent \faIcon{laptop}\textbf{Training and professional development:} Empathy training has been shown to enhance empathetic behaviours, leading to improved interpersonal relationships and team dynamics (see Section \ref{sec:empathy training}). \textcolor{black}{To effectively prepare practitioners for the interpersonal challenges they will encounter in their careers, SE curricula and workplace training programs can incorporate empathy building exercises \cite{gunatilake2025theory}. These exercises can help practitioners develop skills needed to navigate complex interpersonal dynamics, communicate effectively with diverse teams, and manage the emotional aspects of SE. For instance, video-based empathy training can help them improve their conceptual understanding on perspective-taking, allowing them to better understand the viewpoints of their colleagues or users \cite{mitrovic2025video}. Empathy training can also focus on emotional intelligence, conflict resolution, and active listening skills for fostering collaboration and understanding in the workplace. By integrating these elements into training programs, organisations can equip practitioners with the tools they need to create positive and empathetic interactions.} 
 

\section{Method Evaluation and Limitations} \label{sec:limitations}
\textbf{Use of STGT Method:} We evaluate our application of STGT for qualitative data analysis against credibility and rigour, and our outcomes against originality, relevance, and density, as per the STGT evaluation guidelines \cite{hoda2022STGT}. 
\textit{Credibility:} We have provided details on participant recruitment (professional networks, personal contacts, snowballing, and via Prolific), the applied sampling method (convenience, theoretical, and purposive sampling), rationale for all the methodological decisions, data analysis, and the memos. 
\textit{Rigour:} We demonstrate evidence of rigour by presenting examples of sanitised raw data (quotes) throughout Section \ref{sec:findings}. We also illustrate how coding procedures such as open coding and constant comparison were applied to derive codes, concepts, subcategories and categories (see Figure \ref{fig:stgt example}).
\textit{Originality:} In section \ref{sec:introduction} and \ref{sec:related work}, we outline how the limited existing literature motivated this research. We also discuss how our findings complement prior work while offering distinct contributions to the field.
\textit{Relevance:} Our previous studies \cite{gunatilake2023empathy, gunatilake2024enablers, gunatilake2025theory}, along with work by other researchers \cite{cerqueira2023thematic, cerqueira2024empathy}, highlight the growing recognition of empathy as a relevant and important factor in SE contexts.
\textit{Density:} In Section \ref{sec:findings} and in our appendix\footnote{https://doi.org/10.5281/zenodo.17090534}, we have provided evidence of underlying raw data (quotes), that show how the richness and depth of the categories emerged from the data.

\textbf{Limitations:} While our study provides valuable insights into the manifestations of empathy in SE, it is important to acknowledge several limitations. To structure this discussion, we adopted the Total Quality Framework (TQF) for qualitative research \cite{roller2015tqf}, which we found more appropriate for our study than traditional validity frameworks (e.g., internal, external, construct, and conclusion validity), typically used in quantitative research \cite{lenberg2024qualitative}. We present the key limitations aligned with the four components of the TQF: \textit{credibility} (related to data collection), \textit{analysability} (focused on data analysis), \textit{transparency} (regarding reporting), and \textit{usefulness} (concerning the applicability of findings).

The \textit{credibility} component concerns the completeness and accuracy of data collection. A key limitation in this regard is the inherently subjective nature of empathy, which can vary significantly across individuals. We asked practitioners to define empathy without imposing a predefined theoretical lens. While this approach allowed us to capture authentic lived experiences, individual interpretations of empathy may still differ, potentially influencing data consistency. 
Participants for the survey were recruited through Prolific, with compensation provided for their participation. Given that the study focused on empathy, and participants were remunerated, there is a potential for response bias, as participants may feel inclined to provide socially desirable answers. This is a known limitation in survey-based research. However, to mitigate this concern, participation in the study was anonymous, which may have reduced the likelihood of such biases and encouraged more open and sincere responses. In addition, a multi-layered filtering process was applied to assess and refine the quality of participants' responses (see Section \ref{sec:data collection}). 
\textcolor{black}{Another limitation is the potential influence of cultural factors on how participants from diverse backgrounds perceive and report empathy. Empathy may be understood and expressed differently across cultures (Section \ref{sec:Other Factors that Influence Empathy}). These cultural differences could influence both the way participants interpret empathy and how they articulate their experiences during interviews. Further, while our data provides valuable insights into the subjective experiences and perceptions of empathy in SE, the absence of observational data and the potential impact of cultural factors should be considered when interpreting the findings. Our study solely captured verbal and written reports from participants, which may not fully capture the complexities of empathy, especially in real-world contexts. Empathy is a multifaceted construct, and non-verbal expressions, such as body language and facial expressions are key components that were not accounted for in this study. Future studies should incorporate observational data and explore cultural differences more explicitly to offer a more nuanced understanding of how empathy is experienced and expressed across different contexts and cultures.}
\textcolor{black}{In addition, the semi-structured interviews may have been influenced by the interviewer's prompts, and alternative interview questions could have yielded different emphases or additional insights. To address this, we focused on portions of the interviews relevant to empathy manifestations and complemented these with a dedicated survey designed to validate and extend the findings. This triangulation increased confidence in the robustness of our conclusions while acknowledging the inherent limitations of semi-structured interviews.}

The \textit{analysability} component addresses the completeness and accuracy of data analysis \cite{roller2015tqf}. A potential limitation is the inherent subjectivity of qualitative data interpretation. Although we employed a rigorous and iterative analysis process guided by STGT, researcher bias may still have influenced the findings. To mitigate this, the first author's initial coding was reviewed by the co-authors, and the third author, an expert in STGT, peer-reviewed all analytical components. Through multiple rounds of review and discussion, the codebook was collaboratively refined and finalised, enhancing the reliability of the analysis. 

The \textit{transparency} component of TQF emphasises the importance of clear and comprehensive documentation, across all aspects of the research, particularly those related to credibility and analysability. Transparency is supported through the use of thick descriptions, providing rich contextual details that enable readers to assess the transferability of the study's methods, findings, and implications. In this paper, we have thoroughly explained our research design (Section \ref{sec:research design}), findings (Section \ref{sec:findings}), and implications (Section \ref{sec:discussion}), ensuring that readers can critically evaluate and potentially adapt our approach and results to other settings.

The \textit{usefulness} component focuses on the practical value and applicability of the study outcomes.  Our findings provide value to both SE researchers by opening avenues for further investigation, and to practitioners, by offering insights that can inform and enhance day-to-day SE practices.


\section{Conclusion} \label{sec:conclusion}
This study provides a comprehensive understanding of how empathy manifests in SE by exploring its definitions, motivations, useful (or not) SE activities, and influencing factors. Through 22 interviews and a large-scale survey with 116 software practitioners, we identified cognitive empathy, affective empathy, empathic responses (behavioural empathy), and compassionate care as key dimensions of empathy shaping interactions in SE. We also uncovered the motivations driving empathy, the specific SE activities where it is considered useful or not, and the various factors that influence empathy. This study contributes to the growing body of SE research on empathy by offering a more nuanced understanding of its manifestations. Our findings provide actionable insights for practitioners aiming to foster empathy in professional settings and establish a foundation for future research on measuring, fostering, and integrating empathy into SE practices and training programs.

\section*{Acknowledgments}
Gunatilake and Grundy are supported by ARC Laureate Fellowship FL190100035. We express our gratitude to all our participants for generously sharing their experiences, which grounded our research in real-world insights; this work would not have been possible without their contributions.

\appendices

\bibliographystyle{ieeetr}
\bibliography{References}

\end{document}